\documentstyle[12pt,epsfig]{article}  
\newlength{\dinwidth}                       
\newlength{\dinmargin}                      
\def\lsim{\mathrel{\rlap{\lower4pt\hbox{\hskip1pt$\sim$}}
    \raise1pt\hbox{$<$}}}                
\def\gsim{\mathrel{\rlap{\lower4pt\hbox{\hskip1pt$\sim$}}
    \raise1pt\hbox{$>$}}}                
\begin{document}
\begin{flushright}
CERN-TH/96-256\\
DESY 96-194\\
FSU-HEP-960910\\
ITP-SB-96-51\\
INLO-PUB-19/96
\end{flushright}
\vspace*{1cm}
\begin{center}  \begin{Large} \begin{bf}
Deep-inelastic Production of Heavy Quarks\footnote{To appear in the
proceedings of the workshop {\it Future Physics at HERA}, eds. G. Ingelman,
A. De Roeck and R. Klanner, DESY, Hamburg, 1996.}\\
  \end{bf}  \end{Large}
  \vspace*{5mm}
  \begin{large}
E. Laenen$^a$, M. Buza$^b$, B.W. Harris$^c$,  
Y. Matiounine$^d$, R. Migneron$^e$, S. Riemersma$^f$, 
J. Smith$^d$, W.L. van Neerven$^g$\\ 
  \end{large}
\end{center}
$^a$ CERN Theory Division, CH-1211 Geneva 23, Switzerland\\
$^b$ NIKHEF, PO Box 41882, NL-1009 DB Amsterdam, The Netherlands\\
$^c$ Physics Department, Florida State University, Tallahassee, 
FL 32306-3016, USA\\
$^d$ Institute for Theoretical Physics, SUNY Stony Brook, NY 11794, USA\\
$^e$ Dept. of Applied Math., University of Western Ontario, London, Ontario,
  N6A 5B9, Canada
$^f$ DESY IfH-Zeuthen, Platanenallee 6, D-15735 Zeuthen, Germany\\
$^g$ Instituut Lorentz, University of Leiden, PO Box 9506, NL-2300 RA Leiden,
 The Netherlands\\
\begin{quotation}
\noindent
{\bf Abstract:}
Deep-inelastic production of heavy quarks at HERA, especially charm, is an 
excellent signal to measure the gluon distribution in the proton 
at small $x$ values.  By measuring various differential distributions 
of the heavy quarks this reaction permits additional more incisive QCD 
analyses due to the many scales present.  
Furthermore, the relatively small mass of the charm quark,
compared to the typical momentum transfer $Q$, allows 
one to study whether and when to treat this quark as a parton.
This reaction therefore sheds light on some of the most fundamental 
aspects of perturbative QCD.  
We discuss the above issues and review the feasibility
of their experimental investigation in the light of a large integrated
luminosity.
\end{quotation}

\vfill
\begin{flushleft}
September 1996\\
CERN-TH/96-256\\
DESY 96-194\\
FSU-HEP-960910\\
ITP-SB-96-51\\
INLO-PUB-19/96
\end{flushleft}

\section{Introduction}

Since the previous HERA workshop in 1991 
significant progress has been made on the theoretical side 
in understanding the production of heavy quarks
in electron proton collisions. Improvements in available 
experimental techniques and particularly
the expected increase in luminosity amply justify this effort. In general
the progress consists of the calculation of all $O(\alpha_s)$ corrections
to the processes of interest, thus improving the accuracy of the theoretical
predictions both in shape and normalization. At the time of the previous
workshop the only NLO calculations available were for the case of 
inclusive photoproduction \cite{photoincl}. In the meantime 
NLO calculations have also been performed for inclusive electroproduction
\cite{LRSNF,LRSNH,LRSND}, and both have been 
extended to the fully differential cases \cite{FMNR,HS,H}.
Therefore, meaningful and extensive comparisons between theory and 
data can now be made.
In what follows we review how the deeply inelastic 
electroproduction process allows us to explore, in detail, 
three areas of perturbative QCD in particular.

We first discuss the inclusive case, via the structure function
$F_2(x,Q^2,m^2)$. We show that this structure function for the case of 
charm suffers from only very modest theoretical uncertainty,
that its NLO corrections are not too large, and that it is sensitive to
the shape of the small-$x$ gluon density. 
Next we treat single particle differential distributions 
in the charm kinematical
variables, and also charm-anticharm correlations. Because many distributions 
can be studied, many QCD tests can be performed. Examples are tests 
of the production mechanism (boson-gluon fusion), studies of gluon 
radiation patterns, and dependence on scales
such as deep-inelastic momentum transfer $Q$, the heavy quark 
mass $m$ (with enough
luminosity one can detect a sizable sample of bottom quarks), the transverse
momentum of the charm quark, etc. Finally, in the last section, we review the
theoretical status of the boson-gluon fusion description of charm production
at small and very large $Q$. In
essence, it involves answering the question: when is charm a parton?

\section{Structure Functions and Gluon Density}

This section has some overlap with the more detailed review on
heavy flavour structure functions in the structure function section. Here
we only present the most salient features.
The reaction under study is
\begin{equation}
e^-(p_e) + P(p) \rightarrow e^-(p_e') + 
Q(p_1)(\bar{Q}(p_1))+X\,,\label{one}
\end{equation}
where $P(p)$ is a proton with momentum $p$, $Q(p_1)(\bar{Q}(p_1))$ 
is a heavy (anti)-quark with momentum $p_1$ ($p_1^2 = m^2$) and
$X$ is any hadronic state allowed. Its cross section
may be expressed as
\begin{equation}
\label{three}
\frac{d^2\sigma}{dxdQ^2} = 
\frac{2\pi\alpha^2}{x\,Q^4}
\left[ ( 1 + (1-y)^2 )
F_2(x,Q^2,m^2) -y^2
F_L(x,Q^2,m^2) \right]\,,
\end{equation}
where 
\begin{equation}
q = p_e - p_e'\,, \qquad Q^2 = -q^2\,,\qquad x = \frac{Q^2}{2p\cdot q} 
\,, \qquad y = \frac{p\cdot q}{p \cdot p_e}\,. \label{four}
\end{equation}
The inclusive structure functions $F_2$ and $F_L$ were calculated 
to next-to-leading order (NLO)
in Ref.~\cite{LRSNF}. The results can be written as
\begin{eqnarray}
F_{k}(x,Q^2,m^2) &=&
\frac{Q^2 \alpha_s}{4\pi^2 m^2}
\int_x^{z_{\rm max}} \frac{dz}{z}  \Big[ \,e_H^2 f_g(\frac{x}{z},\mu^2)
 c^{(0)}_{k,g} \,\Big] \nonumber \\&&
+\frac{Q^2 \alpha_s^2}{\pi m^2}
\int_x^{z_{\rm max}} \frac{dz}{z}  \Big[ \,e_H^2 f_g(\frac{x}{z},\mu^2)
 (c^{(1)}_{k,g} + \bar c^{(1)}_{k,g} \ln \frac{\mu^2}{m^2}) \nonumber \\ &&
+\sum_{i=q,\bar q} \Big[ e_H^2\,f_i(\frac{x}{z},\mu^2)
 (c^{(1)}_{k,i} + \bar c^{(1)}_{k,i} \ln \frac{\mu^2}{m^2})  
+ e^2_{L,i}\, f_i(\frac{x}{z},\mu^2) d^{(1)}_{k,i}  \, \Big]  \,\Big] \,,
\label{strfns}
\end{eqnarray}
where $k = 2,L$ and the upper boundary on the integration is given by
$z_{\rm max} = Q^2/(Q^2+4m^2)$. The functions $f_i(x,\mu^2)\,, (i=g,q,\bar q)$
denote the parton densities in the proton and $\mu$ stands for the
mass factorization scale,
which has been put equal to the renormalization scale. The 
$c^{(l)}_{k,i}(\eta, \xi)\,,\bar c^{(l)}_{k,i}
(\eta, \xi)\,,
(i=g\,,q\,,\bar q\,;l=0,1)$
and $d^{(l)}_{k,i}(\eta, \xi)$,
$(i=q\,,\bar q\,;l=0,1)$
are coefficient functions and are represented in the
$\overline{\rm MS}$ scheme.
They depend on the scaling variables $\eta$ and $\xi$ defined by
\begin{equation}
\eta = \frac{s}{4m^2} - 1\quad  \qquad \xi = \frac{Q^2}{m^2}\,.
\end{equation}
where $s$ is the square of the c.m. energy of the
virtual photon-parton subprocess
which implies that in (\ref{strfns}) $z=Q^2/(Q^2+s)$. In eq.~(\ref{strfns}) we 
distinguished between the coefficient functions with respect to their origin.
The coefficient functions indicated by
$c^{(l)}_{k,i}(\eta, \xi),\bar c^{(l)}_{k,i}(\eta, \xi)$
originate from the partonic
subprocesses where the virtual photon is coupled to the heavy quark, whereas
the quantity $d^{(l)}_{k,i}(\eta, \xi)$
comes from the subprocess where the virtual
photon interacts with the light quark.
Hence the former are multiplied by the charge squared
of the heavy quark $e_H^2$, and the latter 
by the charge squared of the light quark $e_L^2$ respectively
(both in units of $e$).
Terms proportional to $e_H e_L$ integrate
to zero for the inclusive structure functions.
Furthermore we have isolated the factorization scale dependent logarithm
$\ln(\mu^2/m^2)$. A fast program using fits to the coefficient functions
\cite{RSN} is available.

The first thing to note about eq.~(\ref{strfns}) is that the lowest
order term contains only the gluon density. Light quark densities only
come in at next order, and this is the reason $F_2(x,Q^2,m^2)$ is promising
as a gluon probe.
To judge its use as such,
we must examine some of the characteristics of this observable. These
are: the size of the $O(\alpha_s)$ corrections,
the scale dependence, the mass dependence, its sensitivity
to different gluon densities, and the relative size of the
light quark contribution. These are the issues we investigate in
this section. We take the charm mass 1.5 GeV,
the bottom mass 5 GeV, the factorization scale equal to $\sqrt{Q^2+m^2}$
and choose at NLO the CTEQ4M \cite{cteq4} set of parton densities,
with a two-loop running coupling constant for five flavors and 
$\Lambda = 202 $ MeV, and at LO the corresponding CTEQ4L set, with
a one-loop running coupling with five flavors and $\Lambda = 181 $ MeV. 
\begin{figure}[htbp] 
\hspace*{1.5cm}
\epsfig{file=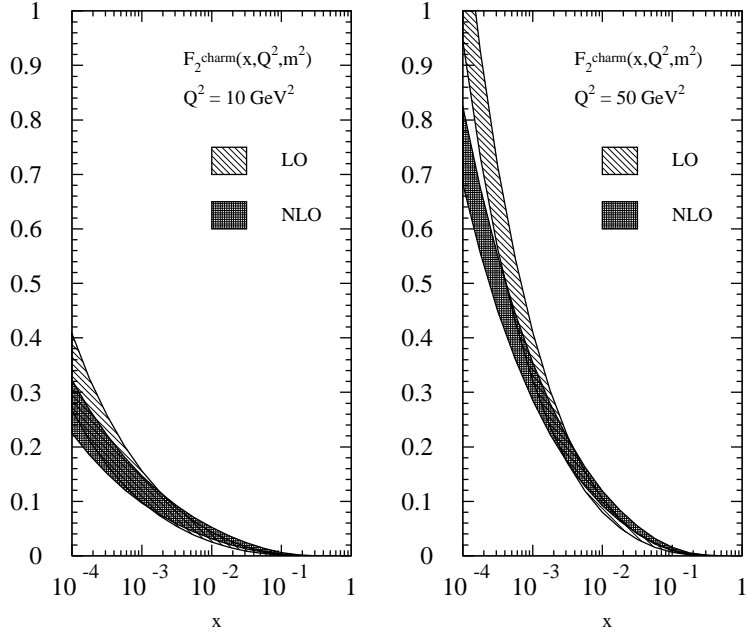,bbllx=0pt,bblly=20pt,bburx=575pt,bbury=880pt,%
        width=10cm,angle=-90}
\caption[junk]{{\it
$F_2(x,Q^2,m^2)$ vs. $x$ at LO and NLO 
for two values of $Q^2$. The shaded areas indicate the uncertainty 
due to varying the charm mass from 1.3 to 1.7 GeV.
  }}
\label{FIGF2C1}
\end{figure}
In Fig.~\ref{FIGF2C1} we display $F_2(x,Q^2,m^2)$ vs. $x$ for
two values of $Q^2$ at LO and NLO.
The scale dependence is much reduced by including
the NLO corrections (when varying $\mu$ from $2$ to $1/2$ times
the default choice, the structure function varies from, 
at LO, at most 20\% and 13\% at
$Q^2 = 10$ and $50$ GeV$^2$ respectively, to at most 5\% and 3\%
at NLO), but the dominant uncertainty is due to the charm 
mass and stays roughly constant, amounting at NLO maximally to about 16\% 
for $Q^2=10$ GeV$^2$
and  10\% for $Q^2=50$ GeV$^2$. The feature that the LO result is
mostly larger than the NLO ones is due to the use
of LO parton densities and one-loop $\alpha_s$,
and scale choice.
Had we used NLO densities and a two-loop $\alpha_s$, or
chosen the scale $\mu$ equal to $m$,
the LO result would have been below the NLO result. In the
first case the size
of the corrections is then about 40\% at the central
values at $Q^2 = 10$ GeV$^2$, and 25\% at $Q^2 = 50$ GeV$^2$,
and in the second case, at small $x$, about 20\%  and 30\% respectively.
\begin{figure}[htbp] 
\hspace*{1.5cm}
\epsfig{file=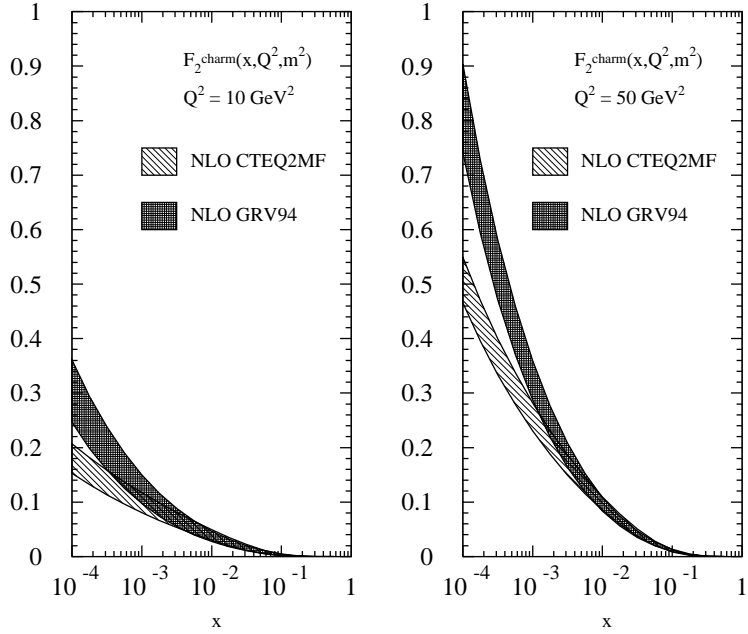,bbllx=0pt,bblly=20pt,bburx=575pt,bbury=880pt,%
        width=10cm,angle=-90}
\caption[junk]{{\it
$F_2(x,Q^2,m^2)$ vs. $x$ at NLO for two choices of parton densities.
The shaded areas again indicate the uncertainty
due to varying the charm mass from 1.3 to 1.7 GeV.
  }}
\label{FIGF2C2}
\end{figure}
In the next figure, Fig.~\ref{FIGF2C2}, we show for the same values of $Q^2$
an important property, namely the
sensitivity of the NLO $F_2$ to different parton density parametrizations.
In this case we compare the CTEQ2MF set \cite{cteq2}, whose gluon 
density stays quite
flat when $x$ becomes small, and the GRV94 set \cite{GRV94}, which
has a steeply rising gluon density. One sees 
that the difference is visible in the structure function.
Finally we remark that the contribution of light quarks to the
charm structure function is typically less than 5\%.
The bottom quark structure function is suppressed by electric charge
and phase space effects and amounts to less than 2\% (5\%)
at $Q^2=10\,(50)$ GeV$^2$ of the charm structure function. Previous
investigations of the scale and parton density dependence of $F_2$
using the same NLO computer codes are available in \cite{vogt}
and \cite{grs}.

We conclude that $F_2(x,Q^2,m^2)$ for charm production is an
excellent probe to infer the gluon density in the proton at
small $x$. 
The NLO theoretical prediction suffers from fairly little
uncertainty, and the QCD corrections are not too large.
See the section on structure
functions in these proceedings for many more details, where also a comparison
with (preliminary) data is shown. Therefore in view of a large 
integrated luminosity, a theoretically well-behaved
observable, and promising initial experimental studies \cite{H1paper,zeus}
a precise measurement at HERA of the gluon density should be possible.

\section{Single Particle Distributions and Heavy Quark Correlations}

In this section we leave the fully inclusive case and examine in more
detail the structure of the final state of the reaction
\begin{equation}
e^-(p_e) + P(p) \rightarrow e^-(p_e') + 
Q(p_1)+\bar{Q}(p_2)+X\,.
\label{react2}
\end{equation}
By studying various differential distributions of the heavy quarks
we can learn more about the dynamics of the 
production process than from the structure 
function alone.

Single particle distributions $dF_2(x,Q^2,m^2,v)/dv$, where 
$v$ is the transverse momentum $p_T$ or rapidity $y$ of the
charm quark, were presented in NLO in
\cite{LRSND} for various choices of $x$ and $Q^2$. The LO distributions
differed significantly from the NLO ones, so that
the effect of $O(\alpha_s)$ corrections on such distributions
cannot be described by a simple K-factor.


The $O(\alpha_s)$ corrections to $F_k(x,Q^2,m^2)$ in a fully
differential form were calculated in Ref.~\cite{HS} using 
the subtraction method.  Recently \cite{H}, these fully differential 
structure functions were incorporated in a Monte-Carlo style program 
resulting in the $O(\alpha_s)$ corrections for reaction (\ref{react2}).
The program for the full cross section, generated 
according to Eq.\ (\ref{three}), allows one to study correlations 
in the lab frame.  The phase space integration is done numerically.  
Therefore,  it is possible to implement experimental cuts.  
It furthermore allows the use of a Peterson type fragmentation function.
For details about the calculational techniques we refer to 
Ref.~\cite{HS,H}.   Here we show mainly results.

\begin{figure}[htbp] 
\hspace*{.5cm}
\epsfig{file=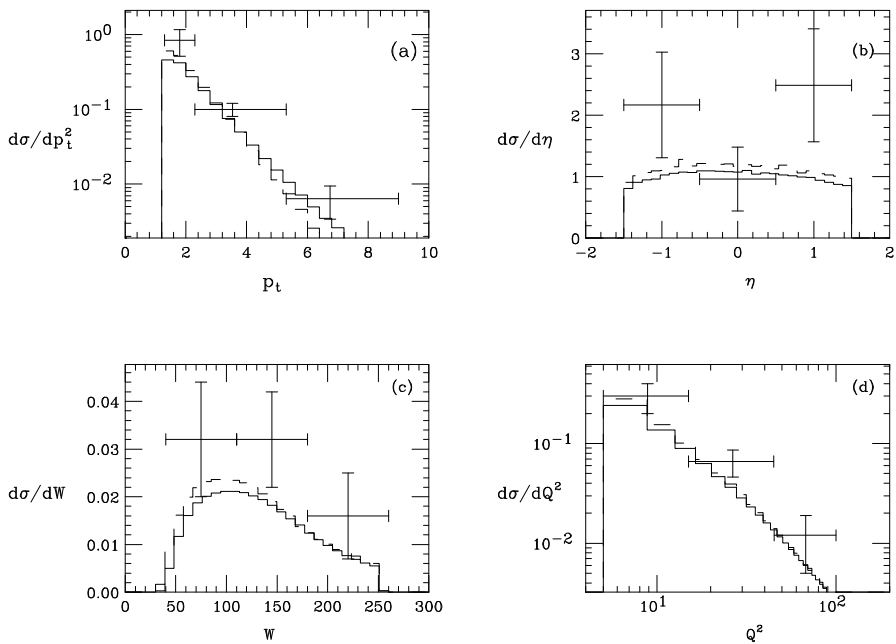,bbllx=0pt,bblly=200pt,bburx=575pt,bbury=600pt,%
        width=13cm,angle=0}
\caption[junk]{{\it
Differential cross sections and ZEUS data. 
  }}
\label{FIGF2C3}
\end{figure}
Shown in Fig.~\ref{FIGF2C3} are various distributions $d\sigma/dv$ 
for the reaction (\ref{react2}), where the heavy (anti)quark
has fragmented into a $D^*$ meson, with $v$ representing
(a) the $D^*$ transverse momentum
$p_T^{{D^*}}$ (b) its pseudorapidity $\eta^{{D^*}}$ (c) the hadronic
final state invariant mass $W$  (d) $Q^2$
for the kinematic range 5 GeV$^2 < Q^2 < $ 100 GeV$^2$,
$0<y<0.7$, $1.3\,{\rm GeV} < p_T^{D^*} < 9 {\rm GeV}$
and $|\eta^{{D^*}}| < 1.5$. The data are from a recent
ZEUS analysis \cite{zeus}. 
The NLO theory curves have been produced by 
using the GRV \cite{GRV94} parton density set, with Peterson
fragmentation \cite{peterson}.
The dashed line is for $\mu=2m$, $m=1.35$ GeV and 
$\epsilon = 0.035$, whereas
the solid line is for $\mu=2\sqrt{Q^2+4m^2}$, $m=1.65$ GeV
and $\epsilon = 0.06$. From Fig.~\ref{FIGF2C3} 
and studies in Ref.~\cite{H1paper}
it is clear that a wide range of 
studies can be and are being performed already at the single
particle inclusive level. Preliminary conclusions \cite{H1paper,zeus}
are that the data follow the shape of the NLO predictions quite well,
but lie above the theory curves. The H1 collaboration
\cite{H1paper} has recently shown clearly
from the $d\ln\sigma/dx_D$ 
distribution that the charm production mechanism is 
indeed boson-gluon fusion,
(after earlier indications from the EMC collaboration \cite{EMC})
as opposed to one where the charm quark is taken from the sea.
Here $x_D=2|\vec{p}_{D^*}|/W$ in the $\gamma^* P$ c.m. frame

Next we examine a few charm-anticharm correlations. 
\begin{figure}[htbp] 
\hspace*{1.5cm}
\epsfig{file=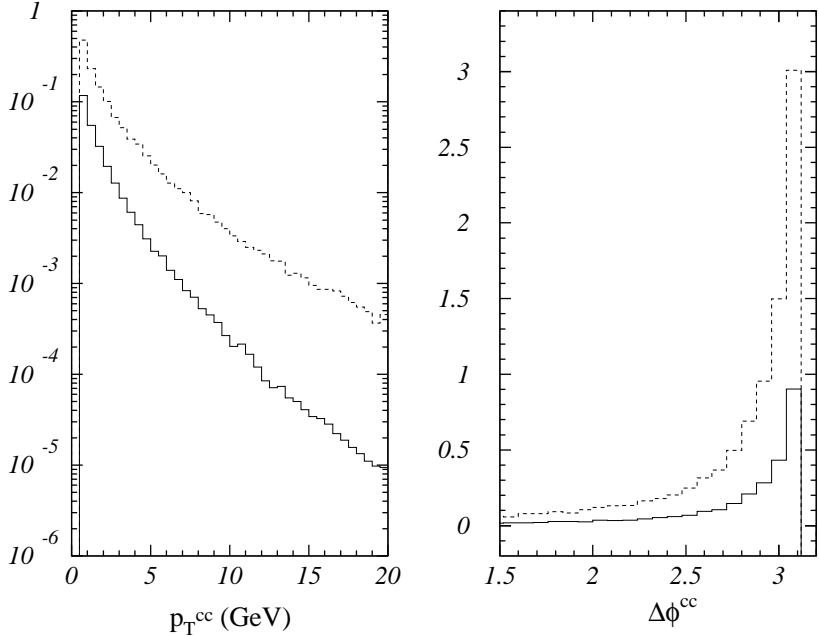,bbllx=0pt,bblly=70pt,bburx=575pt,bbury=800pt,%
        width=10cm,angle=-90}
\caption[junk]{{\it
Differential distributions $dF_2(x,Q^2,m^2,p_{cc})/dp_T^{cc}$ 
and $dF_2(x,Q^2,m^2,p_{cc})/d\Delta\phi^{cc}$
at $x=0.001$ and $Q^2 = 10$ GeV$^2$ (solid) and 100 GeV$^2$ (dashed). 
  }}
\label{FIGF2C4}
\end{figure}
At the
experimental level such correlations are more difficult to
measure since it requires the identification of both heavy
quarks in the final state. However, with the expected large
luminosity that both ZEUS and H1 will collect, such studies
are likely to be done.
As an example we show in Fig.~\ref{FIGF2C4} the $p_T$ distribution 
of the pair, $p_T^{cc}$, and the 
distribution in their azimuthal angle difference, $\Delta\phi^{cc}$
in the $\gamma^* P$ c.m. frame for a particular choice of $x$ and $Q^2$. 
For these figures we used the MRSA$'$ densities \cite{mrsap}.
Both distributions are a measure of the recoiling gluon jet.

In summary, differential distributions of deep-inelastic heavy
quark production offer a rich variety of studies of the 
QCD production mechanism.   Fruitful experimental
studies, even with low statistics,  have been done \cite{H1paper,zeus},
and with a large integrated luminosity we therefore fully expect many more.
We finally point out that besides a LO shower Monte Carlo program
\cite{aroma}, now also a NLO program is available for producing
differential distributions.

\section{When is Charm a Parton?}

We return to the inclusive case to ask the fundamental question in 
the title. The question can be more accurately phrased as follows:
intuitively one expects that at truly large $Q^2$ the charm quark
should be described as a light quark, i.e. as a constituent parton
of the proton, whereas at small $Q^2$ (of order $m^2$) the boson-gluon 
fusion mechanism, in which the charm quark can only be
excited by a hard scattering, is the correct description. This has
been demonstrated recently by H1 \cite{H1paper} and ZEUS in \cite{zeus}. 
In this section we examine where the transition between the two
pictures occurs.

At LO this issue was investigated in \cite{riol}. A picture that
consistently combines both descriptions, the so-called variable
flavor number scheme, is presented and worked out to
LO in \cite{acot}. Here we exhibit where the transition occurs at NLO
\cite{BMMSN}.
In other words we will locate the onset of the 
large $Q^2$ asymptotic region,
where the exact partonic coefficient functions of \cite{LRSNF}
are dominated by large logarithms $\ln(Q^2/m^2)$. These logarithms
are controlled by the renormalization group, and, when resummed,
effectively constitute the charm parton density. 
Here we however restrict ourselves to the onset of the 
asymptotics. Let us be somewhat more precise. In (\ref{strfns})
we can rewrite e.g. all terms proportional to $e_H^2$ as
\begin{equation}
x \int_x^{zmax}\frac{dz}{z}\Big\{
\Sigma(\frac{x}{z},\mu^2) H_{i,q}(z,\frac{Q^2}{m^2},\frac{m^2}{\mu^2})+
G(\frac{x}{z},\mu^2) H_{i,g}(z,\frac{Q^2}{m^2},\frac{m^2}{\mu^2})\Big\}
\end{equation}
where $G(x,\mu^2)$ is the gluon density and 
$\Sigma(x,\mu^2) = \sum_{i=q,\bar{q}}f_{i}(x,\mu^2)$ is the singlet
combination of quark densities.
In the asymptotic regime one may write
\begin{equation}
H^{(k)}_{i,j}(z,\frac{Q^2}{m^2},\frac{m^2}{\mu^2})
= \sum_{l=0}^{k} a_{i,j}^{(k)}(z,\frac{m^2}{\mu^2})
 \ln^l\frac{Q^2}{m^2}\,.
\end{equation}
The effort lies in determining the coefficients $a_{i,j}^{(k)}$. 
Similar expressions hold for the other coefficients in (\ref{strfns}).
Taking the limit of the coefficients in \cite{LRSNF} is extremely
complicated. Rather, a trick \cite{BMMSN} was used, exploiting
the close relationship of the $\ln(Q^2/m^2)$ logarithms with
collinear (mass) singularities. The ingredients are the massless two-loop
coefficient functions of \cite{zn} and certain two-loop operator matrix
elements. 
\begin{figure}[htbp] 
\hspace*{1.5cm}
\epsfig{file=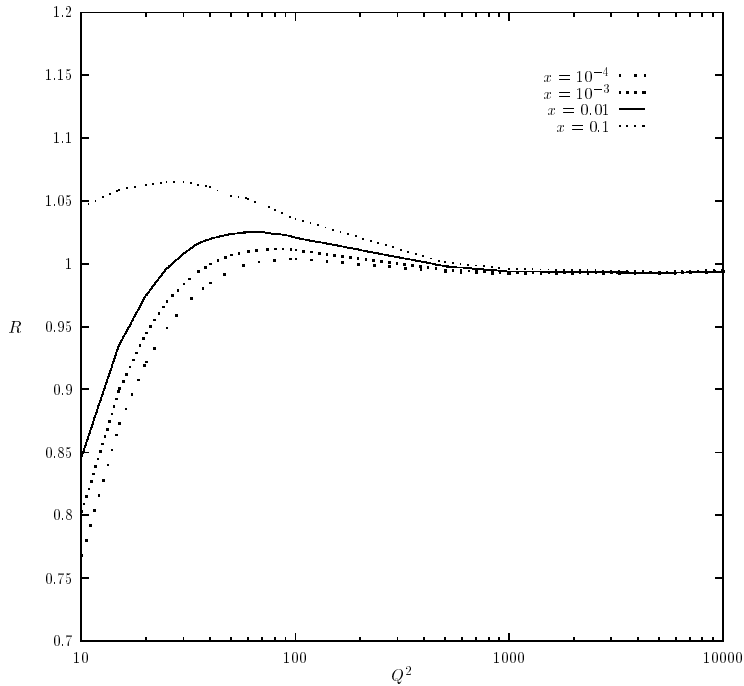,bbllx=0pt,bblly=250pt,bburx=575pt,bbury=700pt,%
        width=12cm,angle=0}
\caption[junk]{{\it
Ratio of the asymptotic to 
exact expressions for $F_2(x,Q^2,m^2)$ for the case of charm.
  }}
\label{FIGF2C5}
\end{figure}
The trick, dubbed ``inverse mass factorization'', essentially amounts
to reinserting into the IR safe massless coefficient functions the
collinear singularities represented by the logarithms $\ln(Q^2/m^2)$.
See \cite{BMMSN} for details.

There is  another advantage to obtaining the asymptotic expresssions.
The terms in eq.~(\ref{strfns}) proportional to $e_L^2$ have been integrated
and full analytical expressions for them exist \cite{BMMSN}, but
in the other terms in eq.~(\ref{strfns}) two integrals still need
to be done numerically. Therefore in the large $Q^2$ region the asymptotic
formula is able to give the same results much faster, as the latter
formula needs no numerical integrations.

In Fig.~\ref{FIGF2C5} we show the ratio of the asymptotic to 
exact expressions for $F_2(x,Q^2,m^2)$ for the case of charm
as a function of $Q^2$ for four different $x$ values.
Here the GRV \cite{GRV94} parton density set was used, for
three light flavors.
We see that, surprisingly, already at $Q^2$ of order 20-30 GeV$^2$ 
the asymptotic formula is practically identical to the exact result, indicating
that at these not so large $Q^2$ values, and for the inclusive structure
function, the charm quark behaves
already very much like a parton.
This is in apparent contradiction with the findings \cite{H1paper},
mentioned in the previous section, that the production
mechanism is boson-gluon fusion, and illustrates that, interestingly, 
the question in the title can have a different answer for inclusive
quantities than for differential distributions having multiple scales.

We finally note that with the results shown in this section 
also the first important step is made for extending
the variable flavour number scheme to NLO.

\section{Conclusions}

In the above we have reviewed the many interesting facets of 
deep-inelastic production of heavy quarks. The possibility
of selecting the heavy quarks among the final state particles
affords a window into the heart of the scattering process,
and allows tests and measurements of some of the most
fundamental aspects of perturbative QCD: the direct determination
of the gluon density, many and varied studies of the heavy
quark production dynamics, and insight into how and when
a heavy quark becomes a parton.


\begin{thebibliography}{99}

\bibitem{photoincl}
R.K. Ellis and P. Nason,
 Nucl. Phys. {\bf B}312 (1989) 551;
J. Smith and W.L. van Neerven, 
Nucl. Phys. {\bf B}374 (1992) 36.

\bibitem{LRSNF}
E. Laenen, S. Riemersma, J. Smith and W.L. van Neerven,
 Nucl. Phys. {\bf B}392 (1993) 162.

\bibitem{LRSNH}
E. Laenen, S. Riemersma, J. Smith and W.L. van Neerven,
 Phys. Lett. {\bf B}291 (1992) 325.

\bibitem{LRSND}
E. Laenen, S. Riemersma, J. Smith and W.L. van Neerven,
 Nucl. Phys. {\bf B}392 (1993) 229.

\bibitem{FMNR}
S. Frixione, M. Mangano, P. Nason and G. Ridolfi, 
 Nucl. Phys. {\bf B}412 (1994) 225.

\bibitem{HS}
B.W. Harris and J. Smith,  
 Nucl. Phys. {\bf B}452 (1995) 109;  Phys. Lett. {\bf B}353 (1995) 535.

\bibitem{H}
B.W. Harris, talk A05057, presented at American Physical Society, Division 
of Particles and Fields 1996 Meeting, Minneapolis, Minnesota, 
10-15 August 1996, to appear in proceedings.

\bibitem{RSN}
S. Riemersma, J. Smith and W.L. van Neerven,
  Phys. Lett. {\bf B}347 (1995) 143.

\bibitem{cteq4}
H.L. Lai et al.(CTEQ Collab.), 
preprint MSUHEP-60426, CTEQ-604, hep-ph/9606399.

\bibitem{cteq2}
H.L. Lai et al.(CTEQ Collab.), Phys. Rev. {\bf D}51 (1995) 4763.

\bibitem{GRV94}
M. Gl\"{u}ck, E. Reya and A. Vogt, Z. Phys. {\bf C}67 (1995) 433.

\bibitem{vogt}
A. Vogt, DESY-96-012.

\bibitem{grs}
M. Gl\"{u}ck, E. Reya and M. Stratmann, Nucl. Phys. {\bf B}422 (1994) 37.

\bibitem{H1paper}
C. Adloff et al. (H1 Collab.), DESY 96-138, hep-ex/9607012.

\bibitem{zeus}
ZEUS Collab., XXVIII Int. Conf. on HEP, Warsaw (1996).

\bibitem{peterson}
C. Peterson, D. Schlatter, I. Schmitt and P. Zerwas, Phys. Rev. {\bf D}27
(1983) 105.

\bibitem{EMC}
J.J. Aubert et al. (EMC Collab.), Nucl. Phys. {\bf B}213 (1983) 31.

\bibitem{mrsap}
A. Martin, R. Roberts and W.J. Stirling, 
Phys. Rev. {\bf D}51 (1995) 4756.

\bibitem{aroma}
G. Ingelman, J. Rathsman and G.A. Schuler, DESY 96-058,
hep-ph/9605285.

\bibitem{riol}
S. Riemersma and F.I. Olness, Phys. Rev. {\bf D}51 (1995) 4746.

\bibitem{acot}
M.A.G. Aivazis, F.I. Olness, and W.-K. Tung, Phys. Rev. {\bf D}50
(1995) 3085; M.A.G. Aivazis, J.C. Collins, F.I. Olness, and W.-K. Tung, 
Phys. Rev. {\bf D}50 (1995) 3102.

\bibitem{BMMSN}
M. Buza, Y. Matiounine, R. Migneron, J. Smith and W.L. van Neerven,
Nucl. Phys. {\bf B}472 (1996) 611;
M. Buza, Y. Matiounine, J. Smith and W.L. van Neerven,
preprint in preparation.

\bibitem{zn}
E. Zijlstra and W.L. van Neerven, 
 Phys. Lett. {\bf B}272 (1991) 127 ; Nucl. Phys. {\bf B}383 (1992) 525.


\end{thebibliography}
\end{document}